\documentclass[twocolumn,secnumarabic,amssymb, nobibnotes, aps, prd]{revtex4-1}
\UseRawInputEncoding

\usepackage{graphicx}
\usepackage{longtable}

\begin{document}
\title{\it{\textbf{Analysis of 2-Body Central Events for $^{129}Xe+^{nat}Sn$ from $8A$ $MeV$ up to $18A$ $MeV$ and for $^{129}Xe+^{197}Au$ at $15A$ $MeV$ and $18A$ $MeV$. }}}

\author{L. Manduci}
\affiliation{Ecole des Applications Militaires de l'Energie Atomique, BP 19 50115, Cherbourg Arm\'ees, France}
%\affiliation{Grand Acc\'el\'erateur National d'Ions Lourds, CEA and CNRS/IN2P3, B.P.~5027, F-14076 Caen Cedex, France}
\affiliation{Normandie Univ, ENSICAEN, UNICAEN,CNRS/IN2P2, LPC Caen, 14400 Caen, France}
%\affiliation{Laboratoire de Physique Corpusculaire, ENSICAEN, Universit\'e de Caen, CNRS/IN2P3, F-14050 Caen Cedex, France} vecchia
\author{O. Lopez, D. Durand, R. Bougault, E. Vient}
\affiliation{Normandie Univ, ENSICAEN, UNICAEN,CNRS/IN2P2, LPC Caen, 14400 Caen, France}
%\author{N. Le Neindre}
\author{M. P\^arlog}
\affiliation{Normandie Univ, ENSICAEN, UNICAEN,CNRS/IN2P2, LPC Caen, 14400 Caen, France}
\affiliation{"Horia Hulubei" National Institute of Physics and Nuclear Engineering (IFIN-HN), RO-077125 Bucharest Magurele, Romania}
%\author{A. Chbihi}
%\author{E. Bonnet}
%\author{J.D. Frankland}
%\author{D. Gruyer}
%\author{P. Marini}
%\affiliation{Grand Acc\'el\'erateur National d'Ions Lourds, CEA and CNRS/IN2P3, B.P.~5027, F-14076 Caen Cedex, France}
%\author{M.F. Rivet}
\author{B. Borderie}
%\author{E. Galichet}
\affiliation{Universit\'e Paris-Saclay, CNRS/IN2P3, IJCLab, 91405 Orsay, France}
%\author{R. Roy}
%\author{J. Moisan}
%\affiliation{Laboratoire de Physique Nucl\'eaire, Universit\'e Laval, Qu\'ebec, Canada G1K 7P4}
%\author{P. Lautesse}
%\affiliation{Institut de Physique Nucl\'eaire, CNRS/IN2P3, Universit\'e Claude Bernard Lyon 1, F-69622 Villeurbanne Cedex, France}
\author{I. Lombardo}
%\author{G. Verde}
\affiliation{INFN Sezione di Catania, via Santa Sofia 64, I-95123 Catania, Italy}
%\author{E. Rosato}
%\affiliation{Dipartimento di Scienze Fisiche and Sezione INFN, Universit\'a di Napoli "Federico II", I-80126 Napoli, Italy}
\author{G. Verde}
\affiliation{Laboratoire des 2 Infinis - Toulouse (L2IT - IN2P3), Université de Toulouse, CNRS, UPS, F-31062 Toulouse Cedex 9 (France)}
\affiliation{INFN Sezione di Catania, via Santa Sofia 64, I-95123 Catania, Italy}
\date{April 2022}
\begin{abstract}
\begin{description}

\item[Background]Study of medium-mass heavy-ion reactions leading to two fragments in the exit channel from barrier to 18A MeV.
\item[Purpose]A special focus is made on fission and quasi-fission for events with two fragments ($Z\geq10$) in the exit channel selected with a specific observable.
\item[Method]Reactions induced by $^{129}Xe$ projectiles on $^{nat}Sn$ and on $^{197}Au$ at energies ranging from
$8A$ MeV to $18A$ MeV were analyzed. Using the fragment ($Z\geq10$) multiplicity equal to 2 and the requirement that the total detected charge to be $90\%$, fission and quasi-fission events were studied for the lowest beam energies using the fission fragment charge distributions, the total kinetic energy distribution (TKE) and its standard deviation $\sigma_{TKE}$.
\item[Results] For the lighter system it is still possible to observe fission events from incomplete fusion. At variance, for the heavier system, Xenon on gold target only quasi-fission is evidenced.
\item[Conclusions] The study of the events characterized by two fragments in the final channel shows that fission, related to fusion in the entrance channel, disappears around 20A MeV for both systems. At lower energies (8A, 12A and 15A MeV) for the Sn target, an evolution with increasing energies towards an asymmetric fission mode is displayed in the fragment charge distributions. This trend can be attributed to the increasing angular momentum as indicated by the out-of-plane angular distribution for light charged particles. The same effect is also observed in the case of the gold target at 15A MeV. However, for this heavier system, a strong memory of the entrance channel leading to quasi-fission is evidenced. A comparison with the Viola systematics, moreover, shows a deviation, greater for the heavier system than for $^{129}Xe+^{nat}Sn$ system. For this latter, at the lower beam energies (8A MeV and 12A MeV) the average Total Kinetic Energy (TKE) shows a behavior close to the Viola systematics, despite the high excitation energies reached in these reactions compared with the ones in previously published studies. The TKE dispersion $\sigma^2_{TKE}$ as well shows the same trend.
A complete understanding of the above results would certainly require precise model calculations at these energies.      
\end{description}  
\end{abstract}
%
%
%\begin{keyword} \sep fusion \sep incomplete fusion \sep DIC \sep asymmetric fission \sep excitation energy  
%\PACS 
%\end{keyword}
%\end{frontmatter}
\maketitle

\section{Introduction}

The present work concentrates on events with two fragments in the exit channel. It continues the study on the reaction mechanisms of the nearly symmetric system  $^{129}Xe + ^{nat}Sn$ at incident energies 8A, 12A, 15A, 18A MeV based on the selection with the variable $E_{Iso}^{Max}$ discussed in reference \cite{PRC_Manduci}.\\ 

The basic assumption made here is the formation of a composite system through a fusion (or incomplete fusion) reaction and its subsequent de-excitation by particles and intermediate mass fragments evaporation or fission for energies lower than 20 A MeV.\\
Following the Bohr hypothesis \cite{bohr}, the two processes can be decoupled : compound nucleus formation and compound nucleus decay once the excitation energy has been distributed to all degrees of freedom. This scenario is observed at low bombarding energies, i.e. close to the barrier. Due to the involved scale, the composite system does not retain any memory of the entrance channel with the exception of the excitation energy and the angular momentum. When it undergoes fission, two fragments are then emitted. These latter may be of nearly equal size. In this case, the macroscopic Liquid Drop Model (LDM) \cite{Wei} accounts for their formation. In the description of the nucleus as a classical and incompressible liquid drop, the repulsive Coulomb interaction and the attractive surface energy of the two fragments create a smooth Potential Energy Surface (PES). Here, the deformed and excited system can just follow one trajectory : the one towards a break up into two equal fragments. However, most of the time these fragments are different in size (asymmetric fission) and the LDM alone can not explain this mass asymmetry. The description of the asymmetric fission needs the inclusion of microscopic effects, such as shell effects or pairing, to consider the influence of the neutron and proton shell closures. These corrections create a PES with different valleys (depending on the entrance energy, the asymmetry and the total mass of the initial system) that the system may follow giving as final state a symmetric or/and an asymmetric fission \cite{karpov}. However there is evidence that shell effects are negligible for higher excitation energies because the PES reverts to a smooth surface with a symmetric fission as in the case of the LDM potential and the compound nucleus will fission mostly symmetrically \cite{andreyev}.\par
For medium-heavy systems ($Z_PZ_T < 1600$, where $Z_P$ and $Z_T$ are, respectively, the projectile and the target atomic number) at energies above the Coulomb barrier, a compound nucleus is easily formed. For heavier systems, instead, the Coulomb repulsion is responsible for the re-separation process with a large probability. In this process, an important nucleonic transfer occurs between the two colliding nuclei and they finally re-separate without forming a compound nucleus. This process is called quasi-fission (QF) and is in competition with the fusion-fission process. It becomes the dominant mechanism for very heavy systems for which the product of the projectile charge multiplied by the target charge exceeds $Z_PZ_T > 1600 $ (\cite{toke}-\cite{hagino}). In this regard, the "orientation effect" may determine the final products of the collision : because of the static deformation of the two nuclei (or of just one of them) there is a variation in the Coulomb barriers whether if the collision occurs as "equatorial-side" (higher Coulomb barrier) or "pole-side"(lower Coulomb barrier) contact; in this case it is enhanced in a collision with a heavy deformed prolate nucleus.   In the first case the reaction may result in symmetric fission while in the second case quasi-fission (QF) sets up giving asymmetric mass distributions \cite{nishio}. \par
For this study, to better understand the role of the system mass, we also studied the heavier system $^{129}Xe + ^{197}Au$ at 15A and 18A MeV.
For the system $^{129}Xe + ^{nat}Sn$ the product of the atomic charges of the projectile ($Z_P$) and the target ($Z_T$) is $Z_PZ_T = 2700$ and for the system $^{129}Xe + ^{197}Au$ is  $Z_PZ_T = 4266$ are larger than the quasi-fission threshold value. This means that these systems should present features typical of the quasi-fission. In the framework of the extra-push model, Bass gave the expression of the fissility effective parameter, $x_{eff}$, which is linked to the entrance channel and is a measure of the competition between the Coulomb interaction to the nuclear potential \cite{bass}, \cite{swiat}. If this parameter is greater than 1 no fusion pocket exists and the compound nucleus cannot therefore be formed. For reactions with the $Sn$ target this parameter is $x_{eff}^{Sn} = 0.898$ while for the $Au$ target its value is $x_{eff}^{Sn} = 1.116$. In the first case the fusion pocket can exists, while in the second one fusion is disfavoured. In a recent study by R. du Rietz \cite{rietz} an average fissility 
parameter ($x_m = 0.75x_{eff}+0.25x_{CN}$ where $x_{eff}$ is the Bass effective fissility parameter and $x_{CN}$ is the fissility compound nucleus for complete fusion) was proposed. For values  $x_m > 0.68$ QF sets on while for values $x_m > 0.765$ QF dominates. Since for $^{129}Xe + ^{nat}Sn$, $x_m^{Sn} = 0.676$ and for $^{129}Xe + ^{197}Au$, $x_m^{Au} = 0.840$,  we expect that quasi-fission dominates for the heavier system. Another criterion, such as the comparison of the entrance channel asymmetry (which is correlated with the fusion probability) for both systems compared with the Businaro-Gallone mass asymmetry \cite{busi}, goes in the same direction : the presence of quasi-fission. \par
In the following, after a short presentation on the experiment and on the INDRA multidetector, we will discuss the observable used to select the events and in the fourth paragraphe we will discuss the events with two fragments in the final state and compare the two systems.

\section{The Experiment}
\begingroup
\squeezetable
\begin{table}
\begin{center}
\begin{tabular}{|c|c|c|c|c|c|}
\hline
$E_{Beam}/A$ $(MeV)$ & $E_{c.m.}$ $(MeV)$ & $E_{c.m.}/V_{C}$&$v_{Lab}$ &
$v_{c.m.}$ & $\Theta_{gr}^{\circ}$   \\ 
\hline
$8$  & $494.6$&1.8&$3.90$&$2.04$&$22.13$   \\
\hline
$12$  &$741.6$&2.7&$4.77$&$2.50$&$12.84$   \\
\hline
$15$  &$926.6$&3.4&$5.32$&$2.79$&$9.79$    \\
\hline
$18$  &$1111.5$&4.0&$5.81$&$3.05$&$7.91$    \\
\hline
$20$  &$1234.6$&4.5&$6.12$&$3.21$&$7.02$   \\
\hline
\end{tabular}
\caption{\label{cin} Kinematic characteristics for the $^{129}Xe+^{nat}Sn$ system at 
different incident energies. The laboratory velocity $v_{Lab}$ and the center mass velocity $v_{c.m.}$ are
in $(cm/ns)$.}
\end{center}
\end{table}
\endgroup
\begingroup
\squeezetable
\begin{table}
\begin{center}
\begin{tabular}{|c|c|c|c|c|c|}
\hline
$E_{Beam/A}$ $(MeV)$ & $E_{c.m.}$ $(MeV)$ & $E_{c.m.}/V_{C}$&$v_{Lab}$ &
$v_{c.m.}$ & $\Theta_{gr}^{\circ}$   \\ 
\hline
$15$  &$1167.3$&2.9&$5.32$&$2.12$&$14.98$    \\
\hline
$18$  &$1400.3$&3.4&$5.81$&$2.33$&$12.02$    \\
\hline
\end{tabular}
\caption{\label{Au} Kinematic characteristics for the $^{129}Xe+^{197}Au$ system at 
different incident energies. The laboratory velocity $v_{Lab}$ and the center mass velocity $v_{c.m.}$ are
in $(cm/ns)$.}
\end{center}
\end{table}
\endgroup
The present study concerns the analysis of the data recorded during the $5^{th}$ INDRA 
campaign for reactions induced by $^{129}Xe$ projectiles on self-supporting (350 $\mu g/ cm^2$ thick) $^{nat}Sn$ targets at different beam energies $E_{Beam}/A$ = 8, 12, 15, 18 MeV with trigger multiplicity $M_{trig} = 1$. We will use also the data from the reactions induced by $^{129}Xe$ on gold target at  E/A = 15, 18 MeV ($M_{trig} = 1$) for comparisons with an heavier and asymmetric system. The apparatus covered angles are from $3^{\circ}$ to $180^{\circ}$.\par 
The experiment was performed at GANIL (Caen, France). The description of the experiment and the
apparatus may be found in references \cite{pouthas1, pouthas2, tabacaru, parlog1, parlog2, tassan, oli}.\par
INDRA can measure ion charge and energy in a wide range and can resolve masses up to 
$Z=4$. Unit charge resolution
was obtained up to $Z=20$ for heavy products stopping in the ion chamber-silicon-CsI telescopes, less than 5 units for $Z=80$.
The energy identification threshold is $\simeq 1$ MeV/nucleon for light fragments and  $1.5-1.7$ MeV/nucleon for fragments of $Z=50$. \par
Tables \ref{cin} and \ref{Au} display the kinematical characteristics of both systems,  $^{129}Xe + ^{nat}Sn$ and $^{129}Xe + ^{197}Au$ respectively, at the different beam energies .  
The Coulomb barrier, calculated with the Bass formula \cite{Bass}, for the Sn target at interaction radius amounts $V_{C} \simeq 275$ MeV whereas for the $^{197}Au$ target it is $V_{C} \simeq 406$ MeV.  As it appears from the 
ratio of the available energy in the centre of mass $E_{c.m.}$ to the Coulomb barrier $E_{c.m.}/V_{C}$, 
in the third column of both tables \ref{cin} and \ref{Au}, the reactions are always produced above Coulomb barrier.

\section{Event selection}
\begin{figure}[!h] 
\begin{center}
\includegraphics*[scale=0.45]{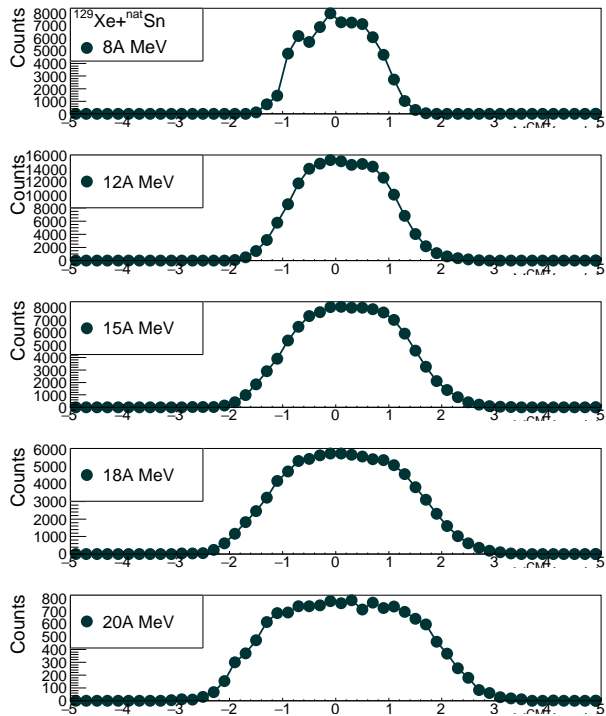}
\caption{Second heavier fragment parallel velocity to the beam direction in the centre of mass frame for incident energies from 8A to 20A MeV for events selected with both conditions : complete events and $E_{Iso}^{Max} \leq 0$ for the system $^{129}Xe+^{nat}Sn$.}
\label{TUTTIbis}
\end{center}
\end{figure}

In order to select events with marked fusion characteristics, the kinematical global observable $E_{Iso}^{Max}$ (\cite{PRC_Manduci}) in term of the velocity components of the heaviest fragment in the event, was used. This observable is defined as : 
\begin{equation}
E_{iso}^{max} = V_{\parallel,max}^{2}-0.5V_{\perp,max}^{2}(1+sin2\phi)\label{ORDI}
\end{equation}
%%%%%%%%%%%%%%%%%%%%%%%%%%%%%%%%%%%%%%%%%%%%%%%%%%%%%%%%%%%%%%%%
where $V_{\parallel,max}$ and $V_{\perp,max}$ are the velocity components of the heaviest fragment in the centre of mass (CM) parallel and
orthogonal to the beam direction and $\phi$ is its azimuthal angle.\\ 
In reference \cite{PRC_Manduci}, the $E_{Iso}^{Max}$ observable was used to select fusion events and to estimate the fusion cross section for each beam energy. 
It was shown that the condition  $E_{Iso}^{Max}\leq 0$ selects one half of the events with marked fusion and incomplete fusion characteristics.\\
The condition $E_{Iso}^{Max}\leq 0 $ alone does not provide a well selected event sample. In fact, when summing up the events for computing the cross section it is reasonable to accept a broader sample constitued also by incomplete events. The above condition on $E_{Iso}^{Max}$ selects also events which had been only partially measured in charge and momentum, i.e. events for which some fragments or particles were not detected. \\
The additional constraint on the total detected charge ensures that the events are well detected. The selected events therefore will satisfy both requirements : "{\it{Complete Events}}  and $E_{Iso}^{Max}\leq 0$ ".\\
By "{\it{Complete Events}} " we mean all the events for which nearly the whole charges produced and their energies were correctly measured. They will be selected by imposing that : \\
\begin{itemize}
\item[1)] the total event charge $Z_{tot}$ (sum of all the charges of the particles and fragments belonging to the event) is larger than $90\%$ of the initial channel. 
\item[2)] the total event pseudo-momentum parallel to the beam direction component :
\begin{equation}%
P_{Tot}= \sum_i Z_iV^{\parallel}_i > 80\%
\end{equation}
where $i$ runs over the detected nuclei.
\end{itemize}
In this way most of the charge and the linear momentum have been correctly measured and only few light charged particles or one intermediate mass fragment (with $Z < 10$) were lost. \\
To support the statement that the selected events are "truly fusion or quasi-fusion events", we display, in figure \ref{TUTTIbis} and for the system $^{129}Xe+^{nat}Sn$, the  parallel velocity (with respect to the beam velocity component) of the second heavier fragment charge for the events selected using both conditions : complete events plus $E_{Iso}^{Max} \leq 0$. The choice for the second heavier fragment, rather than the heaviest one, is due to the fact that this latter is used for the event selection in equation \ref{ORDI} and therefore its velocity distribution will be, by definition, centred around the centre of mass velocity. As one may see in figure \ref{TUTTIbis}, from 8 A MeV till 20A MeV, the centre of mass velocity distributions show a peak, which broadens for increasing beam energies, around zero. For these lower beam energies we can deduce that the fusion-evaporation or fusion-fission could be the dominant mechanism. Already at 18A MeV the peak is very large and at 20A MeV it is rather a plateau. Starting from this last incident energy, if "fusion" and quasi-fusion are still present, the decay mechanism is not any longer evaporation and fission (as already shown in \cite{abdou} and \cite{diego}) but other decay processes which appear to be dominant. This gives rise to forward and backward velocity distributions for the second heavier fragment which at 20 A MeV are still very close. This is why in the study of the events with two fragments in the output channel the highest beam energy considered will be 18A MeV. \\
%Note that the backward component has a lower statistic due to the INDRA efficiency of heavy fragments detection at backward angles.  \par
%
\begin{figure}[!h] 
\begin{center}
\includegraphics*[scale=0.45]{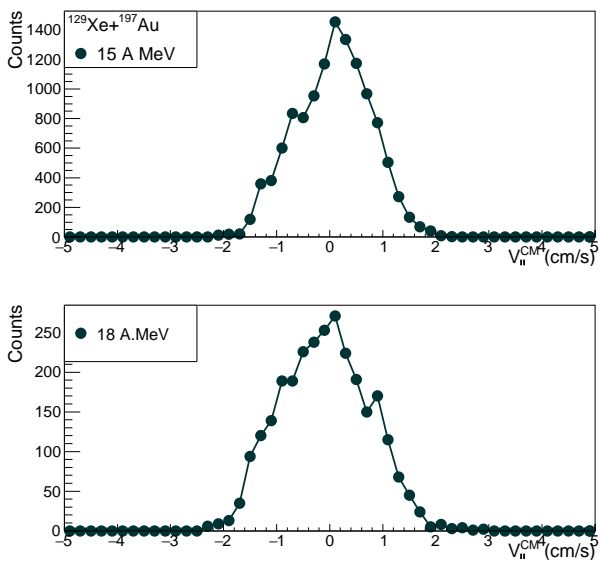}
\caption{Second heavier fragment parallel velocity to the beam direction for Incident energies 15A and 18A MeV for events selected with both conditions : complete events and $E_{Iso}^{Max} \leq 0$ for the system $^{129}Xe+^{197}Au$.}
\label{TUTTI_Au}
\end{center}
\end{figure}
In short, for the $^{129}Xe+^{nat}Sn$ system, both constraints select, with good efficiency and for beam energies lower than 20 A MeV, fusion or incomplete fusion events. \\
Concerning the system $^{129}Xe+^{197}Au$ we applied the same constraints as for the Sn target. Figure \ref{TUTTI_Au} shows, as in the figure \ref{TUTTIbis}, the parallel velocity distributions for 15A MeV and 18A MeV for the second heavier fragment. We observe a velocity distribution centred around the centre of masse without the side contributions seen for the  $^{129}Xe+^{nat}Sn$ at 15A MeV and 18A MeV. 
%
%\begin{figure}[!h] 
%\begin{center}
%\includegraphics*[scale=0.45]{COMPLETE_ZtotPtot_ORDIneg_AnalisiGenerale_Zmax2_VparCM_20-35AMeV.eps}
%\caption{Upper panel : second heavier fragment parallel velocity to the beam direction for beam energies %from 20A to 35A MeV for events selected with both conditions : complete events and $E_{Iso}^{Max} \leq 0$.}
%\label{TUTTIbis2}
%\end{center}
%\end{figure}

\section{Two-fragment multiplicity events}

In this section, we study "fusion" events characterized by two fragments ($Z \geq 10$) in the final state, i. e. the fission of the composite system. Such events were selected requiring, as previously discussed, $E_{Iso}^{Max}\leq 0$ and complete events. \\

Fission has been largely studied in the context of the binary break-up of actinides at low energy.
Most of the theoretical and experimental data were typically produced up to few tens MeV of excitation energy. Experiments performed on spontaneous or thermal neutron induced fission on heavy actinides have provided a large body of data: fission fragment yields and prompt neutron multiplicities which are among the most important variables to elaborate models describing the fission process  \cite{schmidt}, \cite{andreyev}.   \par 
In the present work, fission events were induced in heavy-ion collisions at low and intermediate energies (well above the Coulomb barrier), leading to excitation energies from 1.5 MeV/nucleon up to 4-5 MeV/nucleon and possibly to large internal angular momenta. 
As INDRA can only measure charged particles and provides (with the lowest threshold of about 1 MeV/nucleon) kinetic energies and particle charges, no neutrons spectra are available and no nuclear masses can be identified for the fragments. However, some interesting results can be shown and could trigger phenomenological studies.\par

\subsection{Fission fragment charge, relative velocity and angle distributions, TKE in function of the fragment charge}

\begin{figure}[!h] 
\begin{center}
\includegraphics*[scale=0.44]{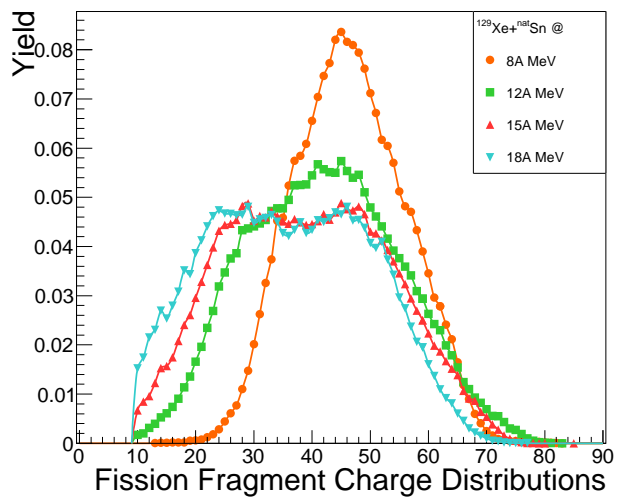}
\caption{Fission fragment charge distributions for beam energies 8A to 18A MeV for the system $^{129}Xe ^{nat}Sn$.}
\label{FFDSn}
\end{center}
\end{figure}

The FFCD (Fission Fragment Charge Distribution) and the TKE (Total Kinetic Energy in the center of mass) are among the most illustrative variables to study fission \cite{schmidt}. \\A global view of fission fragment charge distributions is shown in figure \ref{FFDSn} for the system $^{129}Xe+^{nat}Sn$ at beam energies from 8A to 18A MeV. A smooth transition from symmetric to asymmetric partition is observed. 
%A strong evolution of the mean charge is also evidenced indicating a rapid increase of the excitation energy as beam energy increases.
For the lowest energy at 8A MeV, a typical symmetric fission pattern could be described by the Liquid Drop Model (LDM), since shell effects do not dominate. The fragment distribution is peaked around $Z \simeq 46$. However the distribution is not gaussian : small bumps on both the sides indicate a coexistence with an asymmetric fission mode.  For increasing beam energies, 
%(i.e. increasing excitation energies deposited in the composite system) 
one assists to the transition towards a double humped distribution where the two gaussians are so close to give nearly a plateau distribution. The transition occurs smoothly; in fact, at the beam energy of 12A MeV, one sees that a double asymmetric bump starts to delineate indicating the presence of the asymmetric fission mode. Such a transition is possibly due to the increase of angular momentum transferred to the fissioning systems, as discussed in section IV.3.\par
%The lower panel shows the fragment charge distributions for higher energies : 25A MeV to 35A MeV. Here the fragments are much smaller and no fission patterns are recognizable. In fact, at these high energies, a large amount of light particles and intermediate mass fragments are evaporated thus reducing the total charge of the two final fission fragments.
% 
\begin{figure}[!h] 
\begin{center}
\includegraphics*[scale=0.44]{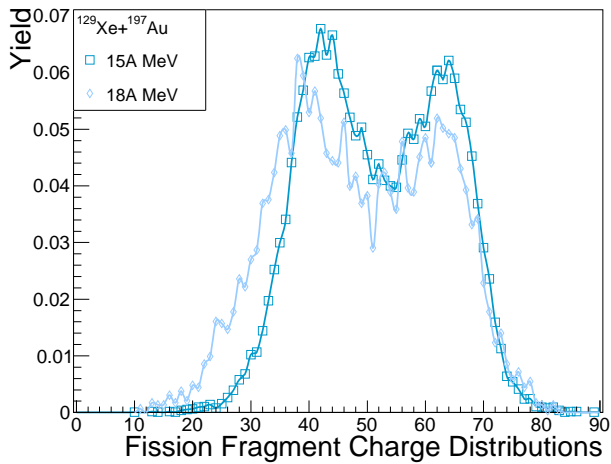}
\caption{Fission fragment charge distributions for beam energies 15A and 18A MeV for the system $^{129}Xe + ^{197}Au$.}
\label{FFDAu}
\end{center}
\end{figure}
Figure \ref{FFDAu} displays the FFCD for the $^{129}Xe + ^{197}Au$ system. Here a clear double humped structure appears which could be originated by fission. However, in the case of such a heavy system, what is observed could be rather quasi-fission with a strong memory of the entrance channel.\par 
%At variance, for the higher energy at 35A MeV, the discussion above is still valid since the fragments are smaller because of the higher energy injected in the system.\\ 
Table \ref{energie} gives the order of magnitude of the mean excitation energy total and per nucleon (evaluated by calorimetry \cite{kaliveda}) for the two systems at the different incident energies. As already mentioned, these values are just a rough estimate because of the uncertainties which could affect \cite{Vient}.\par
%%%
\begingroup
%\squeezetable
\begin{table}
\begin{center}
\begin{tabular}{|c|c|c|c|}
\hline
Target &  $E_{beam}/A$ (MeV)& $<\epsilon>$ MeV/nucleon&$<E^*>$ MeV\\ 
\hline
Sn & 8 & 1.24 & 287\\
\hline
Sn & 12 & 2.22 & 513\\
\hline
Sn & 15 & 2.83 & 652\\
\hline
Sn & 18 & 3.43 & 788\\
\hline
Au & 15 & 2.16 & 633\\
\hline
Au & 18 & 2.63 & 773 \\
\hline
\end{tabular}
\caption{\label{energie}Mean values of the excitation energy (per nucleon and total) for the two systems and different incident energies.}
\end{center}
\end{table}
\endgroup
%Note that, at 15A MeV we approach the excitation energy threshold for the onset of multifragmentation of 3 MeV/nucleon as early suggested \cite{Bizard}, \cite{manduci0}. \par
%
\begin{figure}[!h] 
\begin{center}
\includegraphics*[scale=0.44]{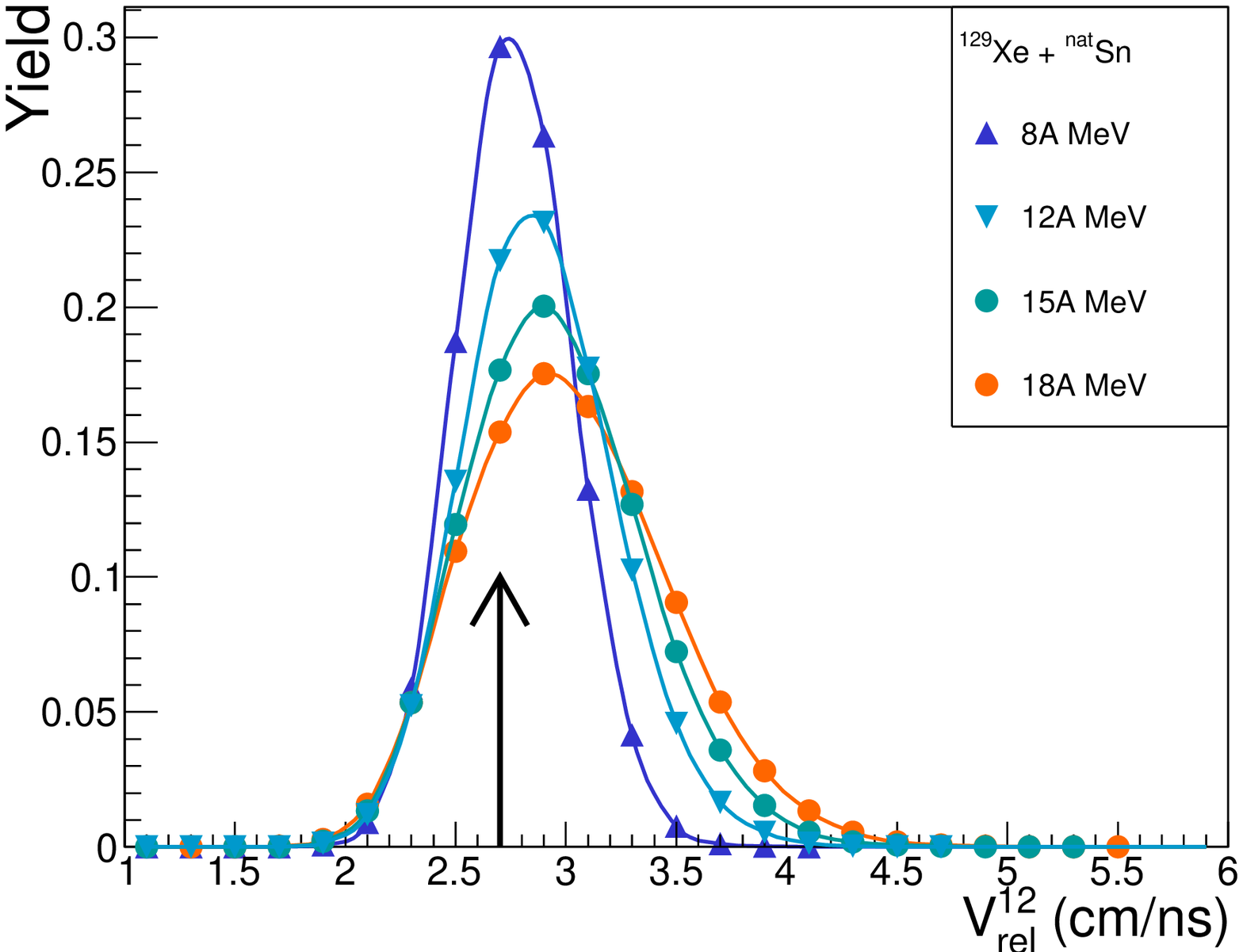}
\includegraphics*[scale=0.44]{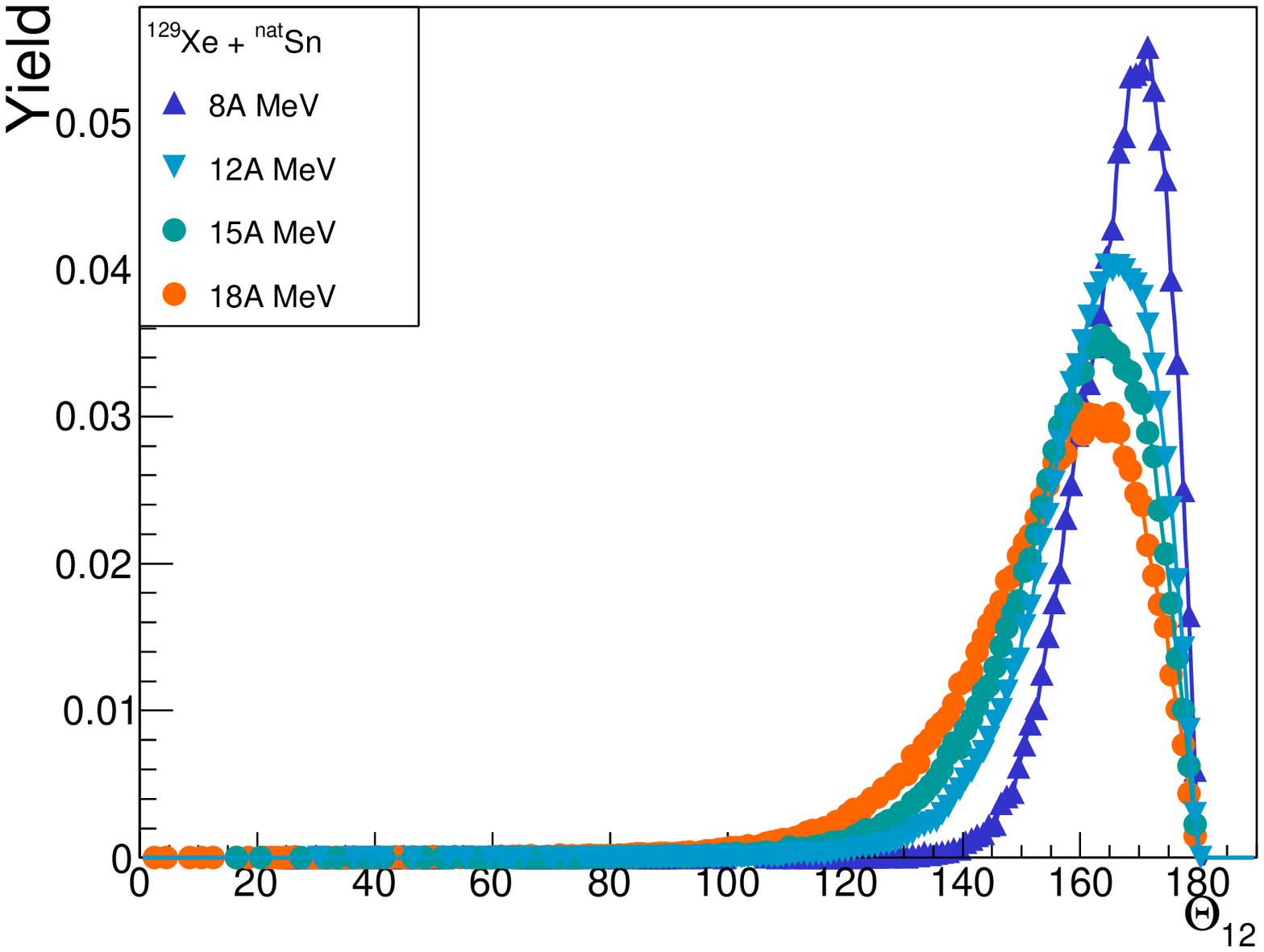}
\caption{Upper panel : Relative velocity distributions between the two fission fragments for beam energies from 8A to 18A MeV. The arrow indicate the Viola velocity. \\ Lower panel : Relative emission angle distributions between the two fission fragments for beam energies from 8A to 18A MeV. The distributions are normalized to the number of events. System $^{129}Xe + ^{119}Sn$. }
\label{V-teta_REL}
\end{center}
\end{figure}
\begin{figure}[!h] 
\begin{center}
\includegraphics*[scale=0.44]{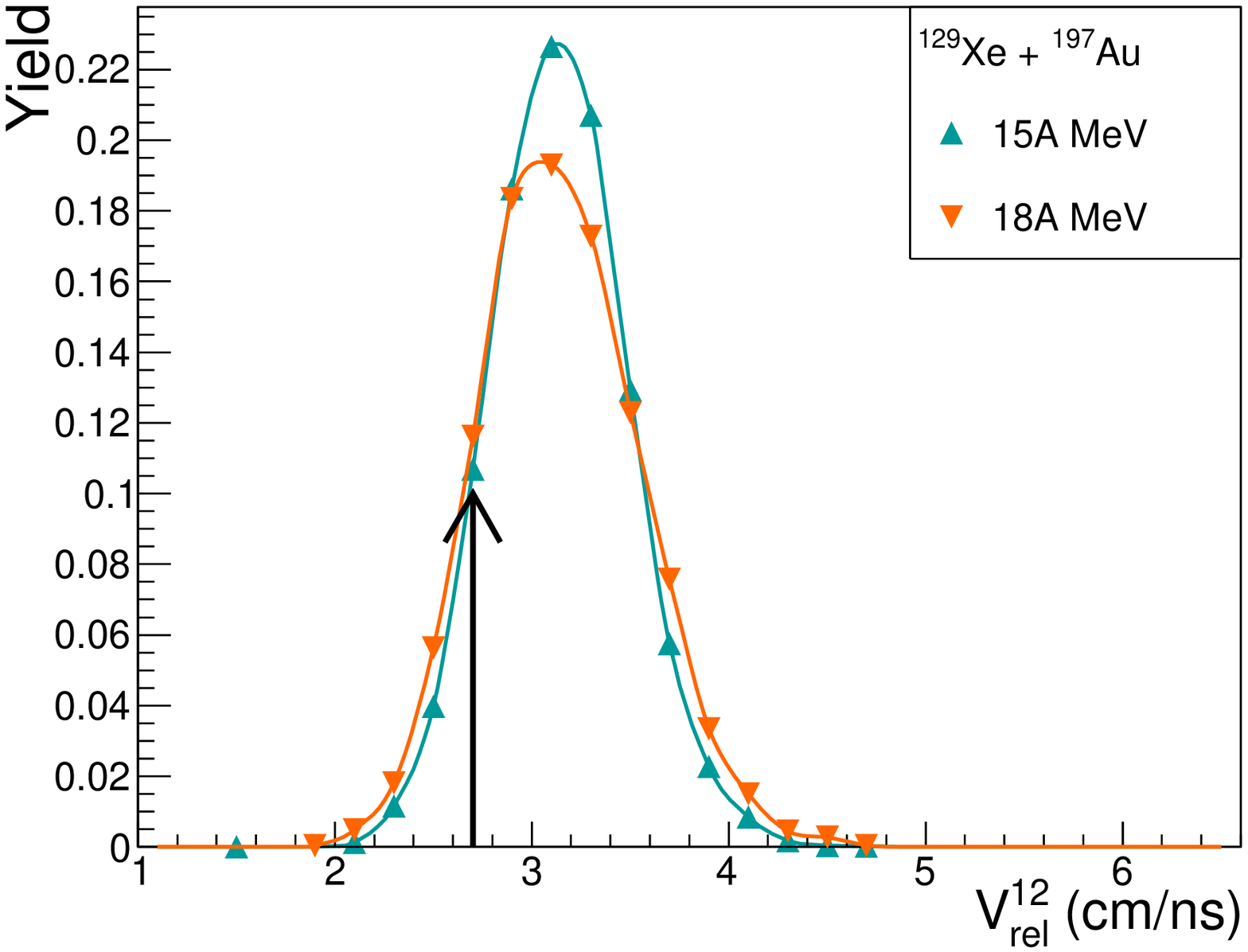}
\includegraphics*[scale=0.44]{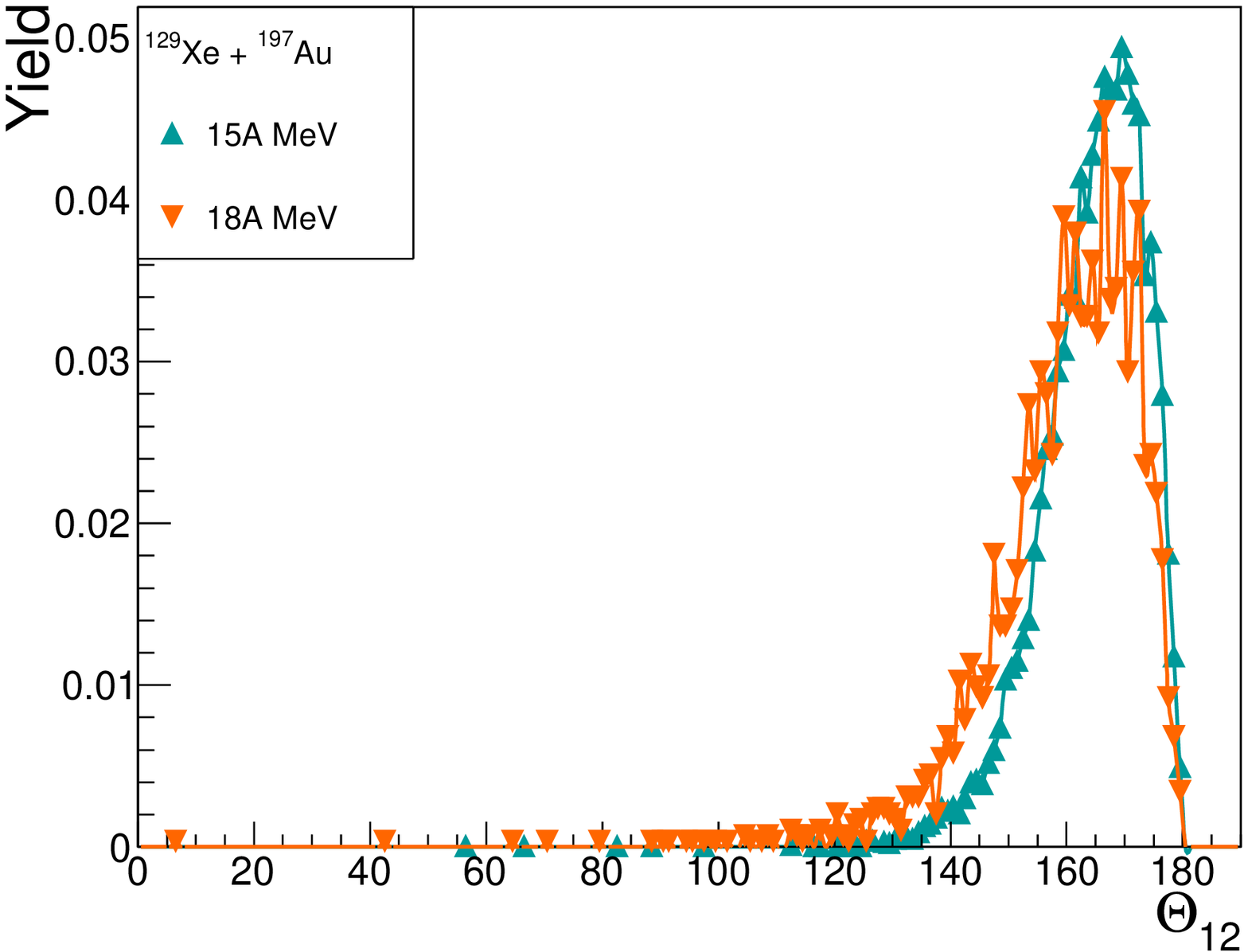}
\caption{Upper panel : Relative velocity distributions between the two fission fragments for beam energies at 15A MeV and 18A MeV for the system $^{129}Xe + ^{197}Au$. The arrow indicate the Viola velocity. \\ Lower panel : Relative emission angle distributions between the two fission fragments at 15A MeV and 18A MeV for $^{129}Xe + ^{197}Au$. The distributions are normalized to the number of events.}
\label{V-teta_RELAu}
\end{center}
\end{figure}
We display in figure \ref{V-teta_REL} the relative two-fragment velocity distributions (upper panel) and the folding angle distributions for the considered incident energies for the system $^{129}Xe + ^{119}Sn$. The arrow shows the Viola systematics value (i.e. $V_{rel}^{Viola} \simeq 2.7 cm/ns$ \cite{viola}) for "fully damped" reaction in the sense of Bohr loss of memory of the entrance channel for the final state. As one may see, for the lowest incident energy, 8A MeV, the mean velocity value is very close to Viola's, while, for increasing beam energies, there is a progressive gaussian broadenning with a sensible displacement towards higher values of the most probable value. This departure from Viola value in the velocity distributions may be understood as a smooth transition with increasing available energy from a dominant mechanism of fusion (or incomplete fusion) with subsequent fission towards a mechanism which is different from pure fission and where the available energy is not completely relaxed so that the system shows a memory of the entrance channel. This could be interpreted as a signature of quasi-fission, where the two nuclei do not form a composite (or compound) system in the very short ($t \simeq 10^{-21}$ s) time in which they enter in contact exchanging few nucleons before re-separating in fission-like fragments. While the presence of quasi-fission, as it will be shown later, is not dominant for the lowest incident energy (8A MeV), it starts to become sizeable as the beam energy increases. \par
%
%There is  and then evolves slowly to larger relative velocities as the energy increases. Here again, the gradual shift is a clear indication of the increasing of the excitation energy which leads to larger fluctuations of the relative velocity distribution and, as such, a deviation from the Viola systematics.
%This can be also an indication of the increasing role of pre-equilibrium processes as beam energy increases leading to an incomplete momentum transfer to the fissioning system.
The lower panel of figure \ref{V-teta_REL} shows the relative emission angle distributions for each incident energy in the center of mass. The relative angle, in the center of mass, is equal or larger than $160^{\circ}$ which means that the fragment were emitted "back-to-back" in a fusion-fission (or quasi-fission) fashion with a full momentum transfer signature \cite{violajr} at the lowest energy. 
%For beam energies from 25A MeV up to 35A MeV a smooth broadening and a displacement of the peak of the distributions toward lower angles is observed, meaning that the fragments were emitted in the forward direction conserving part of the projectile momentum \cite{violajr}.  
%
\begin{figure}[!h] 
\begin{center}
\includegraphics*[scale=0.44]{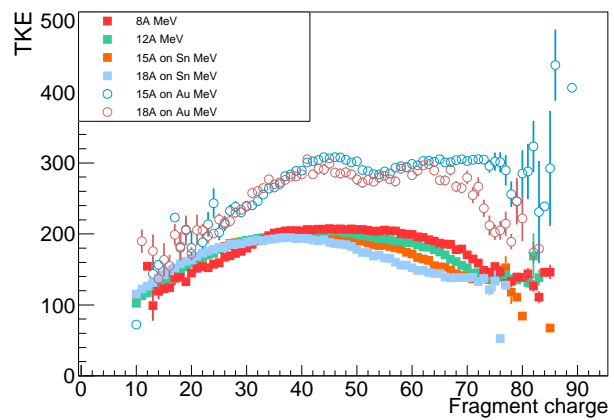}
\caption{TKE (Total Kinetic Energy) profiles in function of the fragment charge for both systems and for all incident energies.  }
\label{profils}
\end{center}
\end{figure}
Figure \ref{V-teta_RELAu} shows the same as figure \ref{V-teta_REL} for the $^{129}Xe + ^{197}Au$ system. In this case the velocity distributions, for both beam energies, depart from the Viola systematics (upper panel) while the folding angle distributions (lower panel) are quite similar to the case of the Sn target. The departure from Viola value in the velocity distributions is more evident than in the Sn target case as expected for heavier systems for which the Coulomb repulsion inhibits the formation of a compound system and so the quasi-fission mechanism is more favoured.
\par
Figure \ref{profils} shows the Total Kinetic Energy (TKE) in the center of mass in function of the fragment charge. The difference in the behavior of the two systems is quite evident : while for $^{129}Xe + ^{119}Sn$, at all incident energies, the TKE values superpose and have nearly the same trend (in particular, at 8A MeV, the trend is very close to the Viola symmetric split values). For $^{129}Xe + ^{197}Au$ (at incident energies as same as the lighter system, i.e. tin target) the TKE values are as higher as roughly 100 MeV, indicating a definite variation in the mechanism of production of the two final fragments. This difference in the TKE value is consistent with the higher mean values of the relative velocity distribution shown in figures \ref{V-teta_REL} and \ref{V-teta_RELAu}, already discussed.

\subsection{Fission Fragment Charge Distributions per $Z_{12}^{Red}$ bins }

To study the role of excitation energy, angular momentum and the competition between symmetric vs asymmetric fission, we define the global variable, $Z_{12}^{Red}$, as follows : 
\begin{equation}
Z_{12}^{Red} = \frac{Z_{12}}{Z_{proj}+Z_{target}}
\end{equation}
where $Z_{12}= Z_1 + Z_2$ is the sum of the fragment charges (where $Z_1$ is the heaviest one).\\
% and  $Z_{proj}= 54$ is the projectile charge and  $Z_{target}$ is $50$ in the case of tin and $79$ for the gold. Therefore, for the $Sn$ target : $Z_{proj}+Z_{target}=104$ and for the $Au$ target : $Z_{proj}+Z_{target}=133$. 
Table \ref{Z12bins} gives the different bins labelled $Tzi$, $i=1,...5$ for the $Z_{12}^{Red}$ reduced sum parameter and the corresponding fragment charge sum for each target. Each bin is 5 units charge wide.\\

\begin{figure}[!h] 
\begin{center}
\includegraphics*[scale=0.46]{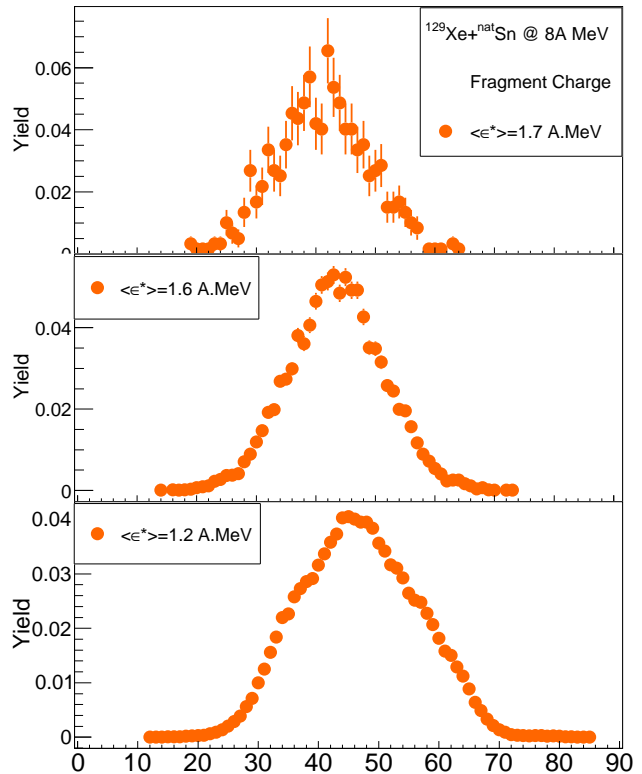}
\caption{Fragment charge distributions for bins of  $Z_{12}^{Red}$ at 8A MeV for $^{129}Xe + ^{119}Sn$. Only the last three bins (Tz3, Tz4, Tz5) are shown because of the low statistics of the first two ones. For each bin is reported the average excitation energy per nucleon. }
\label{Tzi_XeSn_8}
\end{center}
\end{figure}

\begin{figure}[!h] 
\begin{center}
\includegraphics*[scale=0.486]{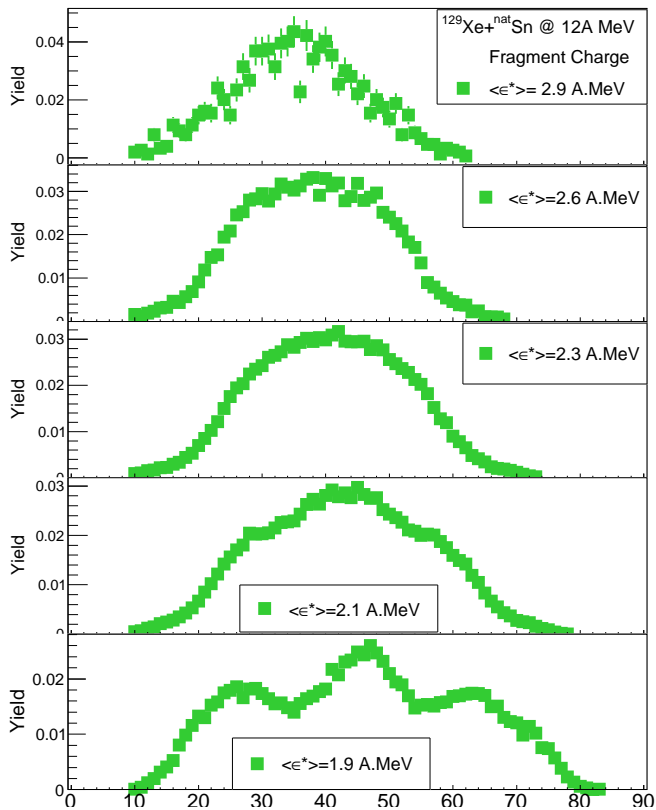}
\caption{Fragment charge distributions for bins of  $Z_{12}^{Red}$ at 12A MeV for $^{129}Xe + ^{119}Sn$. For each bin is reported the average excitation energy per nucleon. }
\label{Tzi_XeSn_12}
\end{center}
\end{figure}

\begin{figure}[!h] 
\begin{center}
\includegraphics*[scale=0.465]{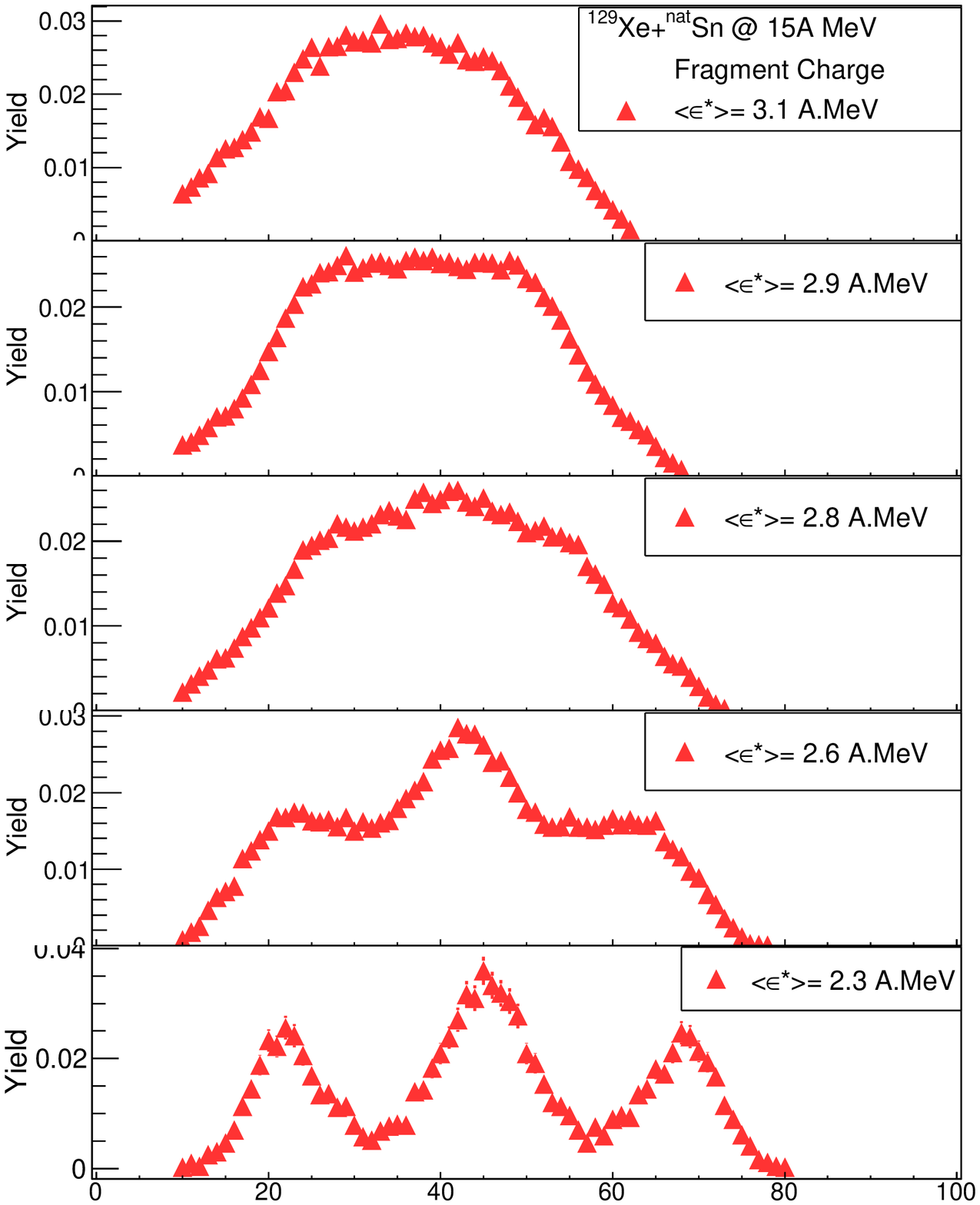}
\includegraphics*[scale=0.465]{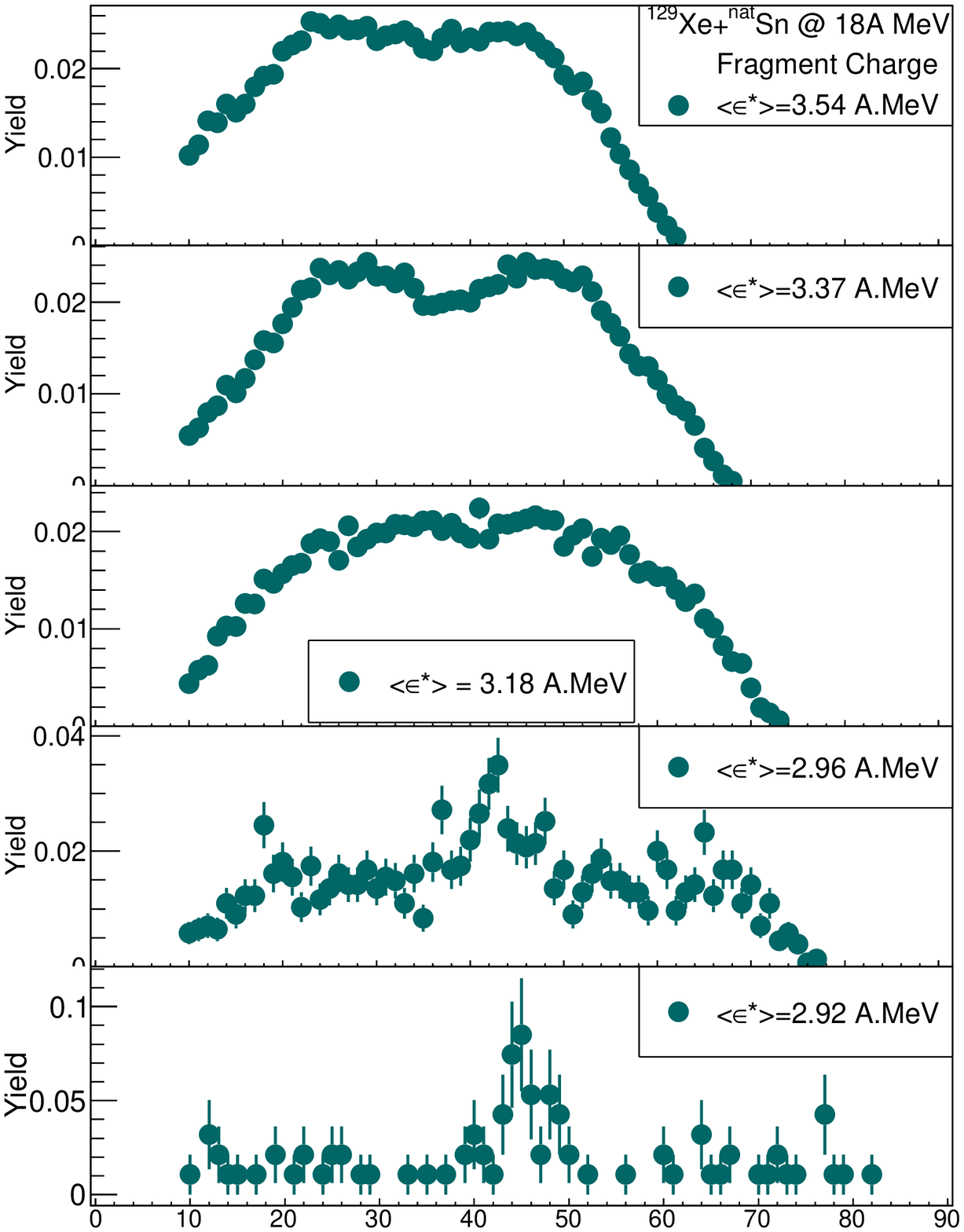}
\caption{Fragment charge distributions for bins of  $Z_{12}^{Red}$ at 15A MeV (upper panel) and at 18A MeV (lower panel) for the system $^{129}Xe+^{nat}Sn$. For each bin is reported the average excitation energy per nucleon.}
\label{Tzi_XeSn}
\end{center}
\end{figure}

\begin{figure}[!h] 
\begin{center}
\includegraphics*[scale=0.46]{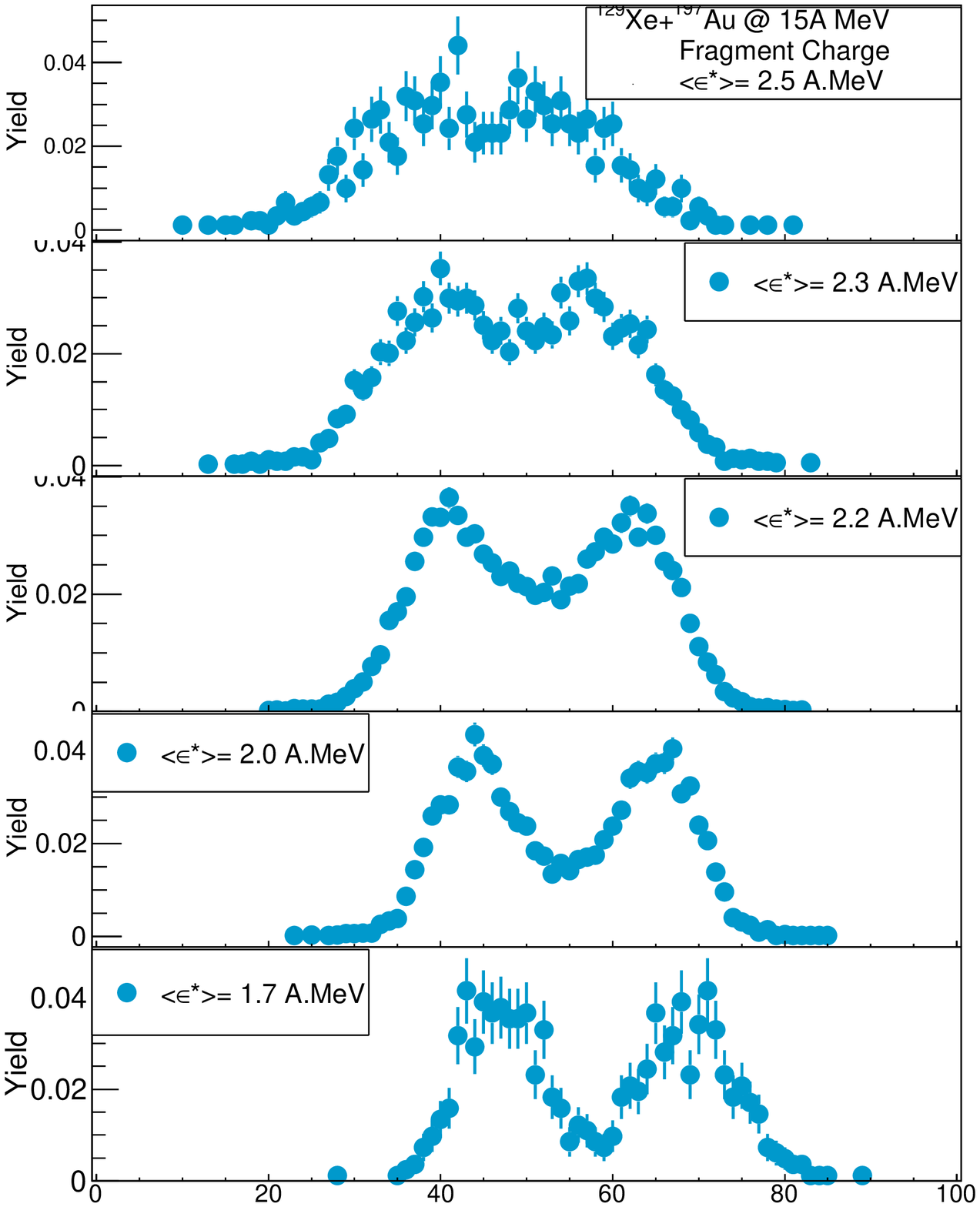}
\includegraphics*[scale=0.46]{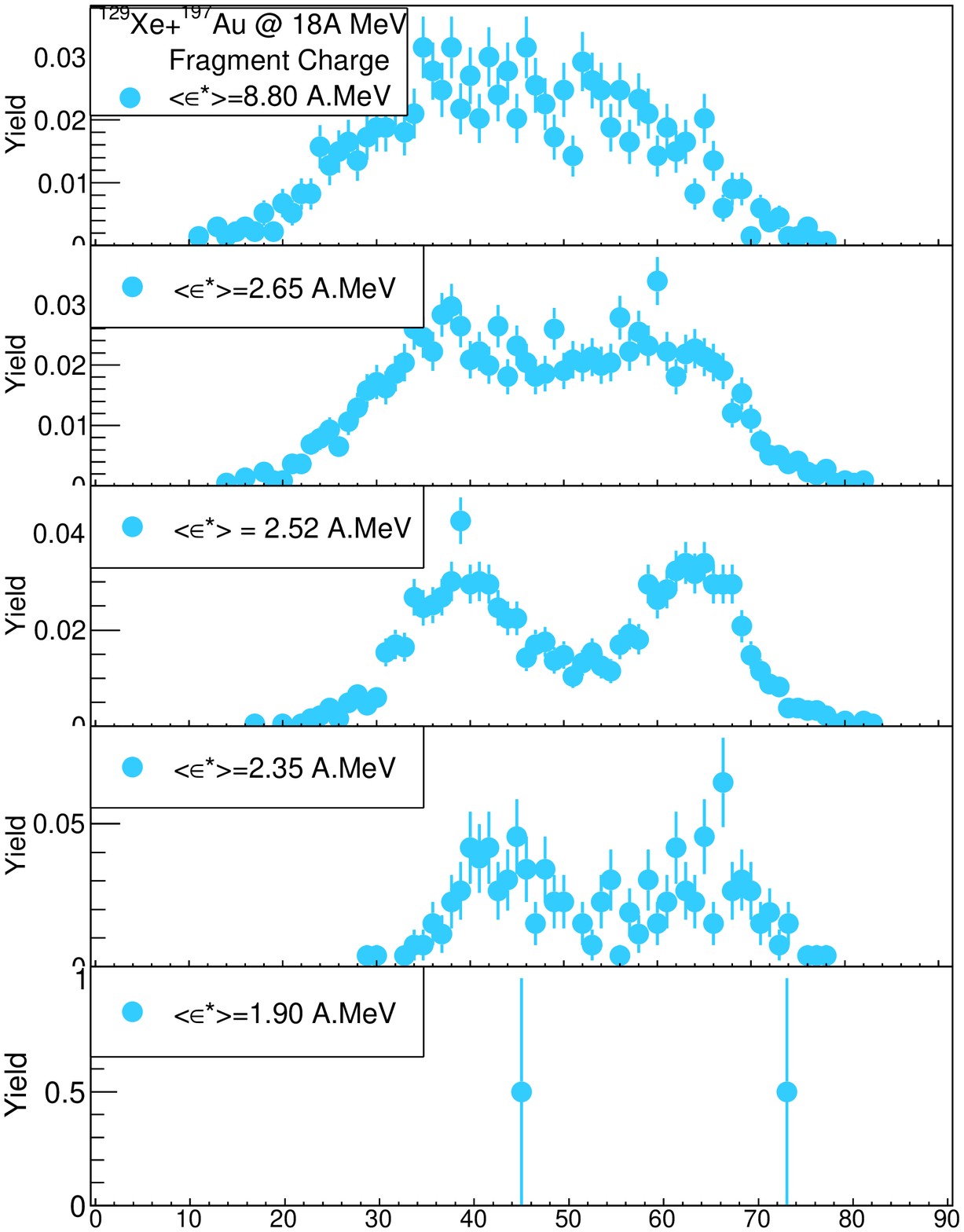}
\caption{Fragment charge distributions for bins of  $Z_{12}^{Red}$ at 15A MeV (upper panel) and at 18A MeV (lower panel) for the system $^{129}Xe+^{197}Au$. For each bin is reported the average excitation energy per nucleon.}
\label{Tzi_XeAu}
\end{center}
\end{figure}

Figures \ref{Tzi_XeSn_8}, \ref{Tzi_XeSn_12}, \ref{Tzi_XeSn} for the Sn target and figure \ref{Tzi_XeAu} for the gold target display, respectively, the evolution of the fission fragment charge distributions for bins of the $Z_{12}^{Red}$ $Tzi$, $i=1,...5$, starting from the top of the figures. For each $Tzi$ bin, are also reported the average excitation energies per nucleon deduced from calorimetry \cite{kaliveda}. Note that their order is decreasing, i.e., the highest value of the excitation energy corresponds to the $Tz1$ bin.\\
The average excitation energies values, altogether with the uncertainties, are listed in the tables \ref{eccibinsSn} and \ref{eccibinsAu} for both systems at the beam energies of interest. 
The assumption that for more central collisions correspond events with larger excitation energy and lower angular momentum transfer is supported by the evolution of the fission fragment charge distributions in the figures, where the mean excitation energy per bin decreases with the increasing values of the charge sum $Z_{12}$. An higher excitation energy is therefore needed to break down the system in two fragments with a smaller size and a larger  emission of light charged particles and intermediate mass fragments. In fact, the light charged particle and intermediate mass fragment multiplicities decrease for decreasing excitation energies as expected. 
\par
\begingroup
%\squeezetable
\begin{table}
\begin{center}
\begin{tabular}{|c|c|c|c|}
\hline
Bins &  $Z_{12}^{Red}$& $Z_{12}^{Sn}$&$Z_{12}^{Au}$\\ 
\hline
Tz1  &  $\leq 0.70$&$\leq 72.8$&$\leq 93.1$ \\
\hline
Tz2  &$(0.70,0.75]$&$(72.8,78]$&$(93.1,99.7]$\\
\hline
Tz3  &$(0.75,0.80]$& $(78,83.2]$&$ (99.7,106.4]$\\
\hline
Tz4  &$(0.80,0.85]$&$(83.2,88.4]$&$(106.4,113]$\\
\hline
Tz5  &$  > 0.85$&$> 88.4$&$> 113$\\
\hline
\end{tabular}
\caption{\label{Z12bins}  $Z_{12}^{Red}$ bins, first column. In the second and third columns are reported the intervals in $Z_{12}$  for the tin and the gold target (event by event), respectively. See text for the variables definition.}
\end{center}
\end{table}
\endgroup
\begingroup
%\squeezetable
\begin{table}
\begin{center}
\begin{tabular}{|c|c|c|c|c|}
\hline
$<\epsilon*>$ & 8A MeV  &  12A MeV  & 15A MeV &   18A MeV \\ 
\hline
Tz1  & -& $2.92 \pm 0.34$ &$3.07 \pm 0.28$&$3.54 \pm 0.31$\\
\hline
Tz2  &$1.95 \pm 0.23$&$2.59 \pm 0.35$&$2.92 \pm 0.29$&$3.37 \pm 0.31$\\
\hline
Tz3  &$1.71 \pm 0.27$&$2.35 \pm 0.30$&$2.76 \pm 0.29$&$3.18 \pm 0.32$ \\
\hline
Tz4  &$1.60 \pm 0.24$&$2.14 \pm 0.27$&$2.56 \pm 0.30$&$2.96 \pm 0.34$\\
\hline
Tz5  &$1.21 \pm 0.27$&$1.91 \pm 0.27$&$2.33 \pm 0.34$&$2.92 \pm 0.28$\\
\hline
\end{tabular}
\caption{\label{eccibinsSn} Mean excitation energy per nucleon (MeV/A) obtained by calorimetry for each bin of the $Z_{12}^{Red}$ variable for the system $^{129}Xe + ^{119}Sn$ at the beam energies 8A, 12A, 15A an 18A MeV.}
\end{center}
\end{table}
\endgroup
\begingroup
%\squeezetable
\begin{table}
\begin{center}
\begin{tabular}{|c|c|c|}
\hline
$<\epsilon*>$  & 15A MeV& 18A MeV\\ 
\hline
Tz1  &$2.48 \pm 0.27$&$2.80 \pm 0.28$\\
\hline
Tz2  &$2.35 \pm 0.28$&$2.65 \pm 0.30$\\
\hline
Tz3  &$2.19 \pm 0.28$&$2.52 \pm 0.30$ \\
\hline
Tz4  &$1.99 \pm 0.29$&$2.35 \pm 0.33$\\
\hline
Tz5  &$1.73 \pm 0.28$&$1.90 \pm 0.18$\\
\hline
\end{tabular}
\caption{\label{eccibinsAu} Mean excitation energy per nucleon (MeV/A) obtained by calorimetry for each bin of the $Z_{12}^{Red}$ variable for $^{129}Xe + ^{197}Au$ at 15A MeV and 18A MeV.}
\end{center}
\end{table}
\endgroup

For the system $^{129}Xe+^{nat}Sn$, in each figure, a slow evolution with the beam energy and with the $Z_{12}^{Red}$ variable is observed.\\
At 8A MeV (figure \ref{Tzi_XeSn_8}), the distributions (which start from the third bin of $Z_{12}^{Red}$ because of the low statistics in the first two bins) show a symmetric fission mode.  In the last bin of $Z_{12}^{Red}$, however, the distribution broadens and departs from a gaussian mode as if an asymmetric fission mode starts to set on. 
%from a two-mode fission at moderate excitation energy and possible larger angular momentum (lower panels) to a symmetric mode at higher excitation energy and lower angular momentum (upper panels). This is particularly evident as the beam energy increases.
At 12A MeV (figure \ref{Tzi_XeSn_12}) the first bin still shows a symmetric fission mode. Then, for increasing values of $Z_{12}^{Red}$, an increase in the asymmetric fission mode is observed, till the last bin in which both modes coexist.\\
At 15A MeV and 18A MeV (figure \ref{Tzi_XeSn}) the symmetric fission mode disappeared and a plateau is observed. For increasing $Z_{12}^{Red}$ values peaks appear which testify the presence of both modes. \\
On the basis of the observation, in figure \ref{V-teta_REL}, of a slight deviation in the mean relative velocity of the two fragments from the Viola systematics, one could wonder if what appears in the last bins at 12A MeV, 15A MeV and 18A MeV could be therefore interpreted as quasi-fission. For example, in the last bin at 15A MeV, the extreme two peaks are, respectively, centered around $Z \simeq 25$ and $ Z\simeq 65$ while the central peak, for symmetric fission, is centered on $Z \simeq 50$. They could have been produced in both mechanisms : asymmetric fission or quasi-fission. \par
%  or if this value of the atomic number is connected to the experimental observation \cite{schmidt} for which the charge of the heavier fragment takes the value of $Z \simeq 54$ (i.e. a mass A centered on $A \simeq 140$ for the heavier fragment in asymmetric fission).
For the heavier system $^{129}Xe + ^{197}Au$ (figure \ref{Tzi_XeAu}) the behavior is different : for each bin of $Z_{12}^{Red}$, at both energies (15A and 18A MeV) two peaks appears starting from the first bin of $Z_{12}^{Red}$ which becomes more well marked in the last bins. These two peaks could rather be quasi-fission for this heavier system. We already saw how the relative mean velocities depart from Viola systematics in a greater extent.\par
The excitation energy decreases from the first to the fifth bin and what changes is the global system size : for higher excitation energies the system evaporated more particles and intermediate mass fragments so that the two fragments, after mass equilibration, have lower mean charge values than the projectile and the target. At variance, in the last Tz5 bin, where the excitation energy is lower,  their mean charge is close to the projectile and the target ones.
This behaviour was already observed, for very low energies, in \cite{Leguillon},\cite{nishio}, \cite{Nishio2}, \cite{Nishio3},\cite{Katsuhisa} where the asymmetric distribution is interpreted as quasi-fission.

\subsection{Light Charged Particle Angular Distributions}

%%%%
\begin{figure}[!h] 
\begin{center}
\includegraphics*[scale=0.48]{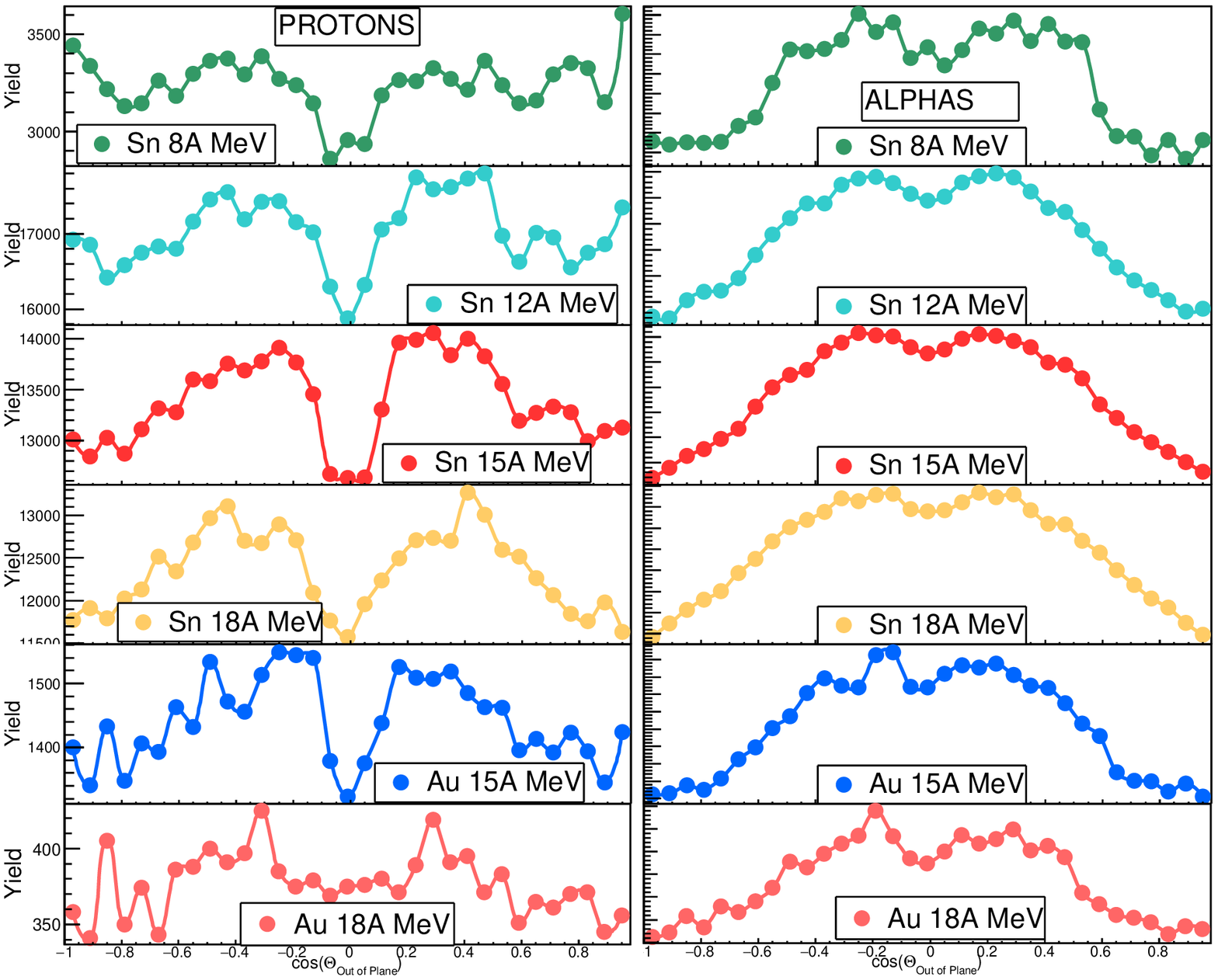}
%\caption{Upper panel : Cosinus of the out-of-plane angle for the system $^{129}Xe+^{nat}Sn$ at 8A, 12A, 15A and 18A MeV. Lower panel : Variance of the same observable.}
\caption{Cosine out-of-plane distributions for protons (left) and alpha (right) for the system $^{129}Xe+^{nat}Sn$ at 8A, 12A, 15A and 18A MeV and for the system $^{129}Xe+^{197}Au$ at 15A MeV and 18A MeV.}
\label{tetaHP_all_en}
\end{center}
\end{figure}
In order to put in evidence the possible role of angular momentum, we present the out-of-plane angular distributions of protons and alpha emitted in coincidence with the two fission fragments. 
First, the reaction plane is defined by the two following vectors: the beam direction and the relative velocity of the two fragments.
%reconstructed centre-of-mass velocity of the two detected fission fragments, i.e the fissioning system.
The out-of-plane angle, $\Theta_{out-of-plane}$, is thus the angle with respect to a vector perpendicular to the reaction plane. The in-plane angle, $\Theta_{in-plane}$, is the angle between the beam direction and the reconstructed center-of-mass velocity. These observables are very sensitive to the rotational energy, i.e. the spin and to the temperature \cite{moretto, colin, steck}.
Figure \ref{tetaHP_all_en} displays the out-of-plane cosine distributions of protons and alpha for 8A, 12A, 15A and 18A MeV incident energies for $Xe+Sn$. For the alphas, the distributions are mostly peaked around $0$ suggesting that most alphas are emitted in the reaction plane as it should be due to angular momentum effect. A small dip around $0$ (i.e. $\Theta_{out-of-plane} =90^{\circ}$) is however observed: this could be due to final state interaction (Coulomb interaction) of the particles with the fission fragments. This 'ejection' process from the reaction plane suggests a rather fast emission and could be used as a clock. For protons, this effect is even more pronounced and the distributions are not anymore peaked around $0$: it is well known that $\alpha$ are much more sensitive to angular momentum than protons. Such findings deserve further analyses based on phenomenological models. 
However, the general evolution of the distributions with beam energies suggest an increase of the transferred angular momentum  with beam energies and probably a shortening of the emission times as expected. 

\subsection{TKE results: comparison with the Viola systematics}

To better compare to low energy data \cite{Itkis}, we study the mean value of TKE as a function of the fissility parameter $Z^2_{CN}/(A_{CN})^{1/3}$, where the charge and the mass of the compound nucleus were estimated as the sum of the charges and the masses (assuming that the ratio $N/Z$ of the fragments is the same as the compound system since masses are not measured) of the two fission fragments, respectively. \\
\begin{figure}[!h] 
\begin{center}
\includegraphics*[scale=0.44]{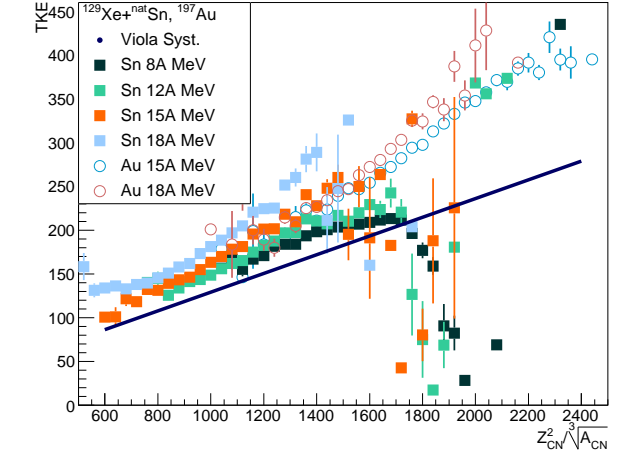}
\includegraphics*[scale=0.44]{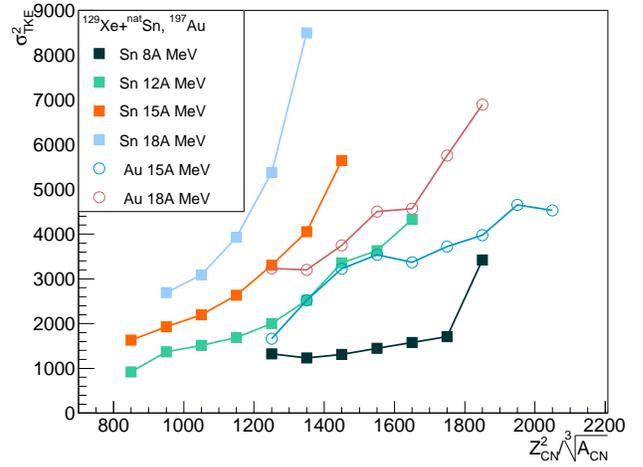}
\caption{Upper panel : average values of TKE as a function of the fissility parameter $Z^2_{CN}/(A_{CN})^{1/3}$ for $^{129}Xe+^{nat}Sn$ at 8A, 12A, 15A and 18A MeV and $^{129}Xe+^{197}Au$ at 15A and 18A MeV. Lower panel : TKE variance $\sigma_{TKE}$ as a function of the fissility parameter for the two systems as explained above.}
\label{sigmaTKE}
\end{center}
\end{figure}
Figure \ref{sigmaTKE} shows in the upper panel the mean TKE values for $^{129}Xe+^{nat}Sn$ at 8A, 12A, 15A and 18A MeV and for $^{129}Xe+^{197}Au$ at 15A and 18A MeV. The line represents the Viola systematics ($0.1071Z^2/A^{1/3} + 22.2$). As one may see, the experimental results are above the Viola systematics as an indication that at those energies we do not find pure fission but a combination of fission and quasi-fission which tends, as the incident energy increases and as the system is heavier, to go more and more towards quasi-fission. 
The closest data to the systematics are those of Xe+Sn at 8A MeV. Slighty above, and starting for lower fissility parameter (i.e. lower composite size), follow those for 12A MeV. Data for the incident energy at 15A MeV for the Sn target overlap to those for 12A MeV till roughly $Z^2_{CN}/(A_{CN})^{1/3} \simeq 1300$ and then set above deviating of about $\Delta(TKE) \simeq 50$ MeV. Finally, always for the Sn target, the data for 18A MeV are very close to both 12A MeV and 15A MeV and they depart for increasing fissility parameters. \\
For $^{129}Xe+^{197}Au$, for both incident energies, starting from $Z^2_{CN}/(A_{CN})^{1/3} \simeq 1500$, the data set definitively above and diverge more and more.\\
Globally, as the incident energy increases and the system is heavier, the departure from the Viola systematics is larger. For a single beam energy one may see that data depart slightly from the Viola line with increasing composite size, especially for the tin target. Due to pre-equilibrium fluctuations the size of the fissioning system  may vary. For lower fissioning system size, the colliding nuclei may reach the potential pocket for fusion and this composite system successively undergoes fission. This would be in agreement with the fact that smaller sizes have greater excitation energy. As the colliding nuclei keep a memory of the entrance channel, the fragments have a larger mass : few nucleons have been exchanged during a very short contact time and then they re-separate. This is quasi-fission marked by 
  a gradual but constant deviation from the Viola line, for both systems but with a more marked trend for gold.\\
Therefore, in the case of the gold target, the behavior is in agreement with the fact that we do not anylonger expect standard fission in the final state but rather quasi-fission for both incident energies. For the Sn target, while for lower energies and, especially for the lowest at 8A MeV, data are closer to the systematics, we may still expect fusion/fission. However, for the higher beam energy the departure from the systematics increases the presence of quasi-fission.
\\ 
The dispersion $\sigma_{TKE}$ is usually linked to the dynamics of the system moving from the saddle to the scission point \cite{Itkis}, \cite{Itkis1}. From the experimental data appears that it is constant for compound nuclei with $Z^2_{CN}/(A_{CN})^{1/3}$ lower than 1000 and for values greater than 1300 it starts to increase linearly. This behaviour is not explained by the Liquid Drop Model (LDM). In the present case one can see that, for the system $^{129}Xe+^{nat}Sn$ at 12A and 15A MeV, the trend is the one predicted while the behaviour of the 8A MeV approach the LDM. However for the system with the gold target the behaviour does not follow neither the systematics nor the other beam energies.

\section{Conclusion}

This work focused on central events for two systems ($^{129}Xe+^{nat}Sn$ and $^{129}Xe+^{197}Au$) at different incident energies from 8A MeV up to 18A MeV. Events with two heavy fragments in the output channel were studied in order to investigate fusion-fission and quasi-fission, in the case of two different systems and for energies exceeding the excitation energies of 100 MeV, typically reached in published studies concerning fusion/fission and quasi-fission at low energy.\\

From FFCD, we saw that, for the quasi-symmetric system $^{129}Xe+^{nat}Sn$ there is an evolution with increasing beam energies from dominant symmetric fission mode at 8A MeV towards more asymmetric fission consistent with the increasing influence of angular momentum as testified by the angular distributions of the associated light charged particles. For larger sizes of the fissioning system (at lower excitation energies) the symmetric and asymmetric fission modes coexists and the asymmetric mode associated to quasi-fission increases as the incident energy increases. On the other hand, for the heavier system $^{129}Xe+^{197}Au$, the FFCD, at both beam energies, show asymmetric fragment distributions suggesting a strong memory of the entrance channel. This fact is confirmed by the subsequent relative fragment velocity distributions in which it is possible to observe a departure from the Viola velocity which increases with the system mass and with increasing beam energies. As already discussed, this was interpreted as dominant fusion-fission at the lowest beam energy (8A MeV), consistent with Viola systematics.
For increasing available energy and pre-equilibrium emission this decay mode is more and more contaminated by quasi-fission decay mode.\\
Therefore,  while for the Sn target there is at least an evolution towards more and more quasi-fission, for the gold target, already at 15A MeV the asymmetric fragment distribution has a memory of the entrance channel, as discussed for the relative fragment velocities. \\
When studying the evolution with the size of the compound undergoing fission, we observe an evolution towards coexistence of both fission modes for the Sn target as the system ($Z_{12}^{Red}$) is heavier, till the emergence of three peaks : asymmetric and symmetric mode altogether. For the gold target, the evolution is towards a definitevely asymmetric (quasi-fission) split.\\
When comparing the Total Kinetic Energy (TKE) in function of the fissility parameter for the both systems with the Viola systematics, one see a clear deviation from the systematics as the incident energy and the size of the fissioning system increase in the case of the tin target. For the gold target system (i.e. the heavier one) this deviation is more evident as the composite system increases in size. For the lighter system (tin target), here again, the lowest energy is the closest to the systematics. Data for increasing beam energies are close to the Viola line only for smaller composite system sizes, where an increasing pre-equilibrium left a fissioning system such that it could partially fuse (reaching the potential pocket) and then fission. The progression of the departure from the systematics is therefore two fold : in beam energies and in system size.\\ 
 The behaviour of the $\sigma^{2}_{TKE}$ dispersion shows a trend similar to previous studies reported in literature for the Xe+Sn system (especially for 8A MeV), while for the Xe+Au system a completely different behaviour is observed.\\
We are not aware of fission modelisation at such high energies as studied here. Fission dynamics involved in the case of heavy ion reactions at these energies are different from the studied dynamics close to the Coulomb barrier. To deeply understand the behaviour discussed above, it would be useful to develop systematic microscopic model calculations to be compared with our results.

%%%%%%%%%%%%%%%%%%%%%%%%%%%%%%%%%%%%%%%%%%%%%%%%%%%%%%%%%%%%%%%%%%%%%%%%%%%%%%%%%%%%%%%%

\end{document}